\documentclass[10pt,prb,twocolumn,showpacs]{revtex4}
\usepackage{graphicx}
\usepackage{epsf}
\usepackage{ifthen}
\usepackage{amsmath}
\usepackage{amssymb}
\usepackage{subfigure}
\usepackage{psfrag}
\usepackage{float}
\floatstyle{plaintop}
\restylefloat{table}

\usepackage{subeqnarray}
\begin{document}
\title{AA-stacked bilayer square ice between graphene layers?}

\author{S. Fernandez Mario, M. Neek-Amal\footnote{Corresponding author: neekamal@srttu.edu},  and F. M. Peeters}

\affiliation{Universiteit Antwerpen, Department of Physics,
Groenenborgerlaan 171, BE-2020 Antwerpen, Belgium.}

\date{\today}

\begin{abstract}

Water confined between two layers with separation  of a few Angstrom forms layered two-dimensional ice structure. Using large scale 
 molecular dynamics simulations with the adoptable ReaxFF interatomic potential we found that flat monolayer ice with a rhombic-square structure
nucleates between graphene layers which is non-polar and
non-ferroelectric. Two layers of water are found to
crystallize
 into a square lattice close to the experimental found AA-stacking [G. Algara-Siller et al.
Nature \textbf{519}, 443445 (2015)]. Each layer
has a net dipole moment which are in opposite direction. Bilayer ice is also non-polar and non-ferroelectric. For three layer ice
we found that each layer has a crystal structure similar to
monolayer ice.

\end{abstract}

\maketitle

\emph{Introduction}. The phase diagram of water and its
extraordinary properties have been an  interesting topic of research
in biology, chemistry, and physics for many decades. Depending on
the hydrophobic confinement width several two-dimensional ice
structures can be formed~\cite{prl2003,jchem2003,PNAS,ACCOUNT}.
Different theoretical methods, e.g. molecular dynamics (MD)
simulations  using different force fields~\cite{prl2003,ACCOUNT},
density functional theory~\cite{arxiv2015}, and Monte Carlo
simulations~\cite{PCCP}, have been used to study ice formation in
the presence of high pressure. In particular, monolayer ice was
proposed by Zangi and Mark~\cite{prl2003,jchem2003} using MD
simulations by applying a five site and tetrahedrally coordinated model,
i.e. TIP5P. They confined water between two parallel plates and applied a high lateral pressure ($P_l$) of about 1\,GPa and found a
non-flat monolayer of ice. 

Recently, it was found
experimentally that confined water exists as a quasi two-dimensional
layer with different properties than those of bulk
water~\cite{apl2015,nat2015}. Graphene, the two-dimensional
allotrope of carbon~\cite{Novoselov}, was used in a recent
experiment to  confine water~\cite{nat2015} into monolayer, bilayer and three layers. Using transmission electron microscopy(TEM) square lattice structures was observed.
 The lateral
pressure for confining water between two sheets of graphene can be
estimated~\cite{nat2015,prl2012,natnano} to be about 1\,GPa using
the van der Waals (vdW) adhesive energy between two layers which is
typically around 20 meV\AA$^{-2}$. This experiment was supported by
MD simulations that showed that by increasing the pressure, bilayer square
ice (three layer ice) with a lattice constant of 2.82\AA~nucleates where the graphene layers are separated by a distance
h=9\AA~(11.5\AA)~\cite{nat2015}. However, the MD simulations failed to reproduce the experimental found AA-stacking of bilayer ice. Ab-initio calculations found  that
monolayer ice confined between hydrophobic graphene layers can be
rippled or flat, depending on the confinement width and lateral
pressure~\cite{arxiv2015}. However, this DFT study is based on a small
supercell which therefore  missed structures that involve more than
4 water molecules.


Using the reactive bond order potential we reveal
new physics of confined ice between two graphene layers.  We performed annealing MD
simulations starting from high temperature, i.e. 400\,K and found the low temperature
minimum energy configuration, and determined the structure of
monolayer, bilayer and three layer ice. We evaluate the different
energy terms, charge distribution, and hydrogen bond strength of
confined water. The ReaxFF potential takes into account the
polarization of charge within the molecules, it makes our study very
different from all previous investigations. The studied systems
are found to be all flat, non-polar and non-ferroelectric where the
microscopic structure depends on the number of ice layers. We found for the first time a flat structure for confined ice and estimated the vdW energy
between two ice layers.

\begin{figure*}
\begin{center}
\includegraphics[width=0.53\linewidth]{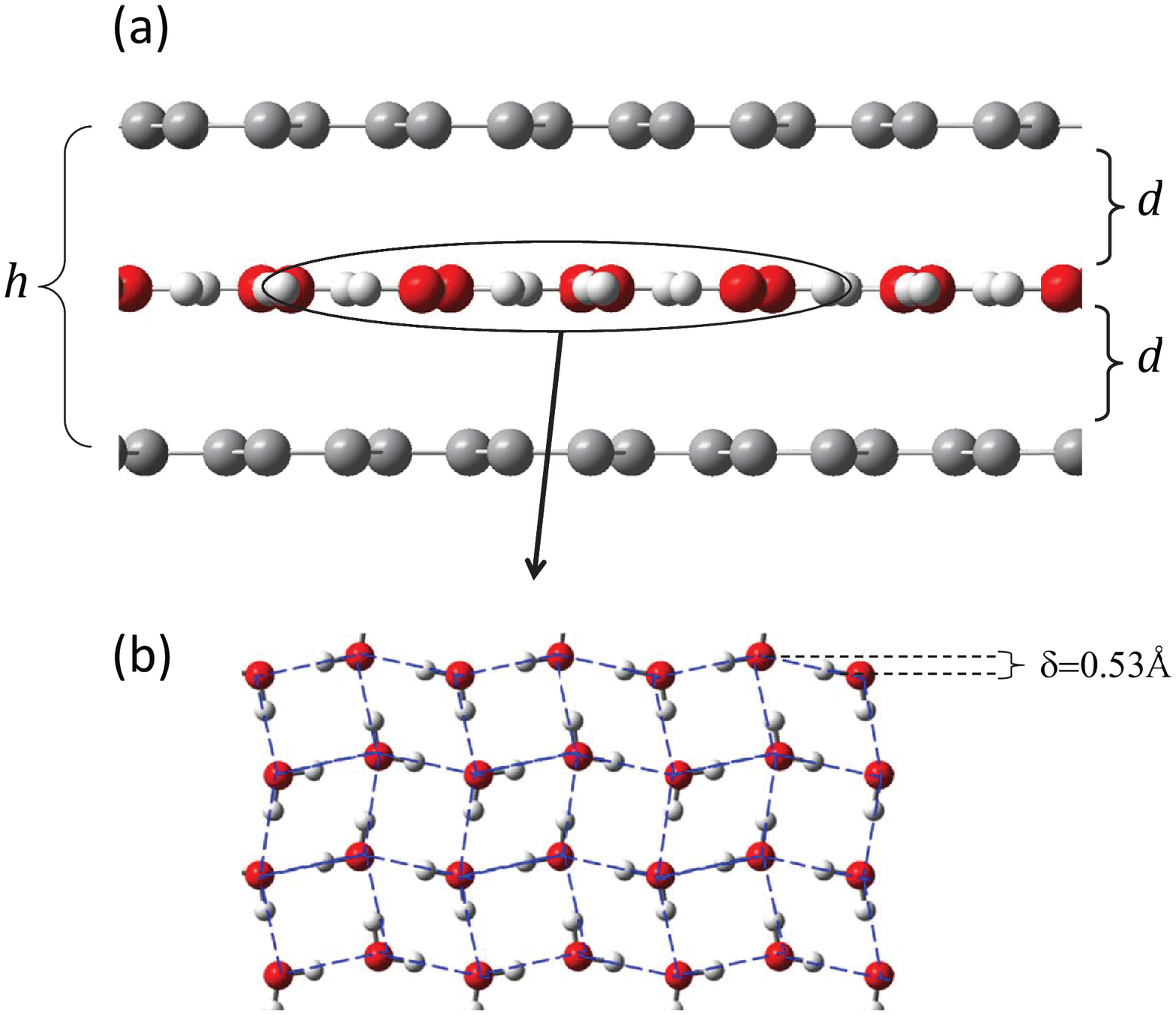}
\includegraphics[width=0.35\linewidth]{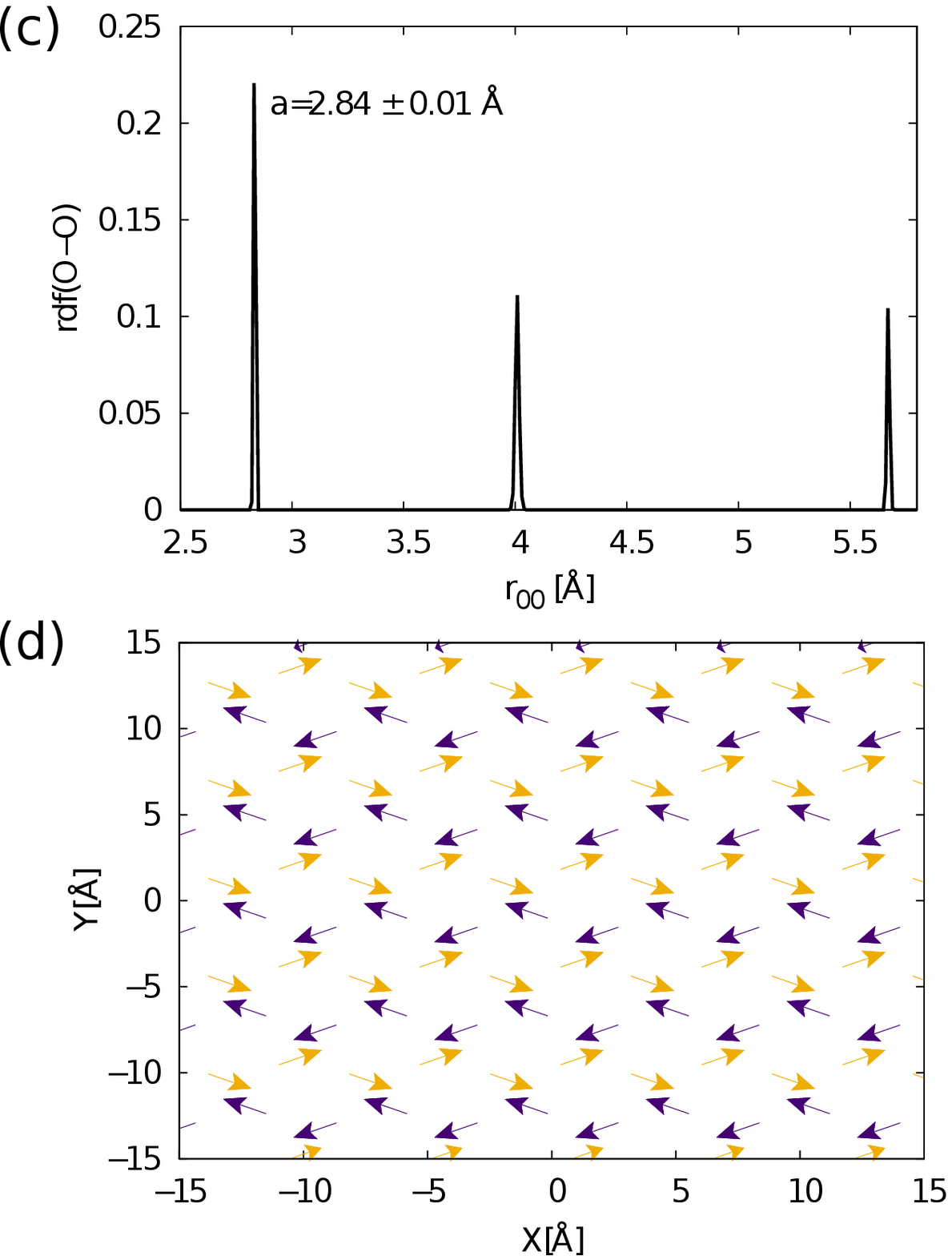}
\caption{\label{fig1} (color online) The top (a) and side (b) view of
the relaxed monolayer of ice between two graphene layers (red(grey) circles indicate O(H) atoms.  The
corresponding (c) radial distribution function for O-O distances and
(d) local dipoles. }
\end{center}
\end{figure*}
\emph{Method and Model}. We employed molecular dynamics (MD)
simulations
 using reactive force fields (ReaxFF~\cite{reax}) potentials in the well-known large-scale
 atomic/molecular massively parallel simulator ``LAMMPS"~\cite{lammps}. ReaxFF potential accounts
 for possible bond-formation and bond-dissociation of different bond orders.
 It also contains Coulomb and van der Waals potentials to describe non-bond interactions between all atoms.
 One of the main advantages of ReaxFF is that it calculates the polarization
 of charge within the molecules which is achieved by using electronegativity and hardness parameters based on
 the electronegativity equalization method and charge equilibration (QEq) methods. Therefore, we believe that ReaxFF is a
 better candidate to simulate  water and the corresponding interaction between water and graphene.
 Furthermore, the ReaxFF potential allows bond extension/contraction in water as well as angle bending and it allows
 charge relaxation over each atom. This is in contrast to the traditional force fields for water,
 e.g. SPC and TIP4P~\cite{tip2005} (a rigid planar four-site interaction potential for water) that keep
 the water molecules rigid during MD simulations.

 The computational unit cell contains 34848
 carbon atoms and  17100 $\times$n water molecules in the system where $n$ is the number of ice layers.
Before performing minimization we do an annealing MD simulation by
performing a NPT simulation starting at 400\,K and ending at 0\,K,
in order to find the true simulation box size and O-O distances. The starting
high temperature guarantees that the O atoms and H-bonds can find their
minimum energy configurations during the very slow annealing process.
Then, the total energy is minimized using the iterative conjugate
gradient (CG) scheme.

\emph{Monolayer of ice}. We start from a random distribution for the
H-bonds of the water molecules which are distributed in a dense
square structure, with O-O distance equal to 2.8\AA,~between two
graphene layers.  The initial O-O distance is set only to have the experimental observed density for confined ice but during annealing which we start from 400\,K, the positions are allowed to change. The graphene layers are rigid and separated by a
fixed distance of 6.5\,\AA~having AB-stacking. The results for the
water layer are independent of the exact stacking configuration of
the graphene layers. By minimizing the potential energy, we found a
flat rhombic-square lattice structure, see Fig.~\ref{fig1}. The side
view of the minimum energy configuration is shown in
Fig.~\ref{fig1}(a) which is a flat monolayer of ice with successive
arrangements of square and rhombic building blocks, see
Fig.~\ref{fig1}(b). If the O atoms remains in the same plane, the H
atoms should also be in that plane in order to preserve the
symmetry. The other possibility would be a buckled (puckered)
structure which results in non-flat ice~\cite{prl2003} (which was
not found with our simulations)~\cite{Note1}. Our results are
partially in agreement with ab-initio results where Corsetti et
al~\cite{arxiv2015} used non-local vdW exchange   correlations and
scanned both the confinement size and lateral pressure. However
their unit cell (called Ab/Cd) was too small, i.e. it
contained only 4 water molecules,
  in order to find the aforementioned asymmetry effect.
  Our found structure for monolayer
ice is in agreement with the MD simulation results using the SPC/E
  model~\cite{nat2015}.  In fact the TEM results of the recent
experiment~\cite{nat2015} is not a perfect square lattice
layer~\cite{private}.

The crystalline structure and lattice constant of monolayer ice
can be determined from the radial distribution function (RDF), see
Fig.~\ref{fig1}(c). We found the O-O distance in flat ice to be
a=2.84$\pm$0.01\AA. The obtained lattice constant is in good
agreement with the experimental value of a=2.81$\pm$0.02\AA. The
angle `H--O--H' is found to be $\theta$=106.31$\pm$0.03$^o$ and is
identical for all ``H--O--H' bonds. The H-bonding energy is about
-0.16\,eV/water (-15.43\,kJmol$^{-1}$) for each water molecule which is
in the range of the H-bond energy of bulk ice\,~\cite{chaplin}, i.e. -(13-32)
kJmol$^{-1}$. It is worth to mentioning that the vdW energy stored
in the system is $E_{vdW}$ = 2.43\,eV/atom which is positive and the
Coulomb energy is $E_{coulomb}=$-0.56\,eV/atom which is negative.
These numbers show that in dense ice the O atoms repel each other
strongly (note that $a=r_{OO}\approx$~2.8\AA~lies in the repulsion
region of the vdW energy function~\cite{prbSTM}). In Table
I, we list all relevant quantities  for the different studied systems.

\begin{figure}[h!]
\centering
\includegraphics[width=\linewidth]{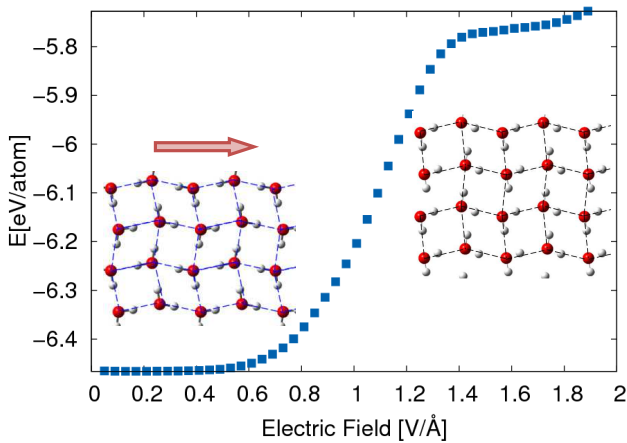}
\caption{(color online) Energy of an ice monolayer relaxed
between two graphene layers when applying an in-plane electric field
along the direction of the arrow (left inset). Note the change of the H-bond
orientation (right inset). } \label{fig:Electricfield}
\end{figure}

The corresponding local dipoles of the water molecules in the
minimum energy configuration is shown by the arrows in
Fig.~\ref{fig1}(d). Interestingly, the net dipole is zero and the
system is non-ferroelectric which is in agreement with ab-initio
results~\cite{arxiv2015} and is in disagreement with the TIP5P model
prediction~\cite{ferro}. We were able to deform the H-bonding orientations using an in-plane electric
field of about 1\,V/$\AA$ (see Fig.
\ref{fig:Electricfield}). As shown in Fig.
\ref{fig:Electricfield}, by increasing the electric field (along the direction of the shown
arrow in the inset of Fig.~\ref{fig:Electricfield}) from zero
to around $|E|\simeq1$ V/$\AA$ we find a structural transition
from the minimum energy configuration with
zero dipole moment to a new higher energy configuration with a net
non-zero dipole moment, i.e. the local dipoles of the water molecules
rotate in the direction of the electric field (see right inset of Fig.~2).

\begin{figure}[b]
\includegraphics[width=\linewidth]{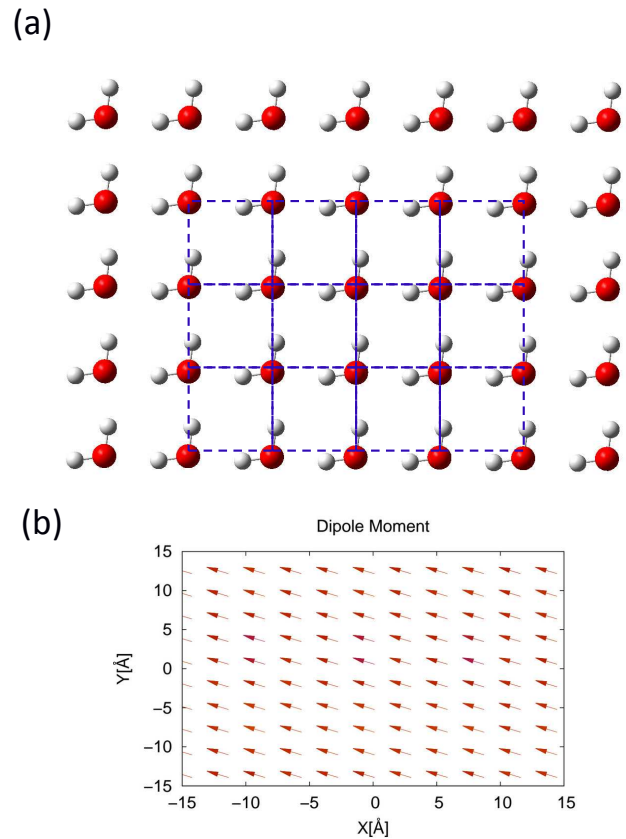}
\caption{\label{figBeta} (color online)(a) A polar lattice of
monolayer ice confined between two graphene layers which has about 10\,meV/atom higher energy than the non-polar structure. (b) The
corresponding local dipoles.}
\end{figure}

For non-polar flat monolayer ice there are many possible structures.
We made a detailed investigation of them and present the results of one of the 
most relevant ones. We performed
additional MD annealing and minimization by starting from an initial
configuration where the arranged H-bonds are in three dimensions and
two H-bonds of each water molecule have the same orientation as the
equivalent bonds in all the others. We found that the minimum energy
configuration is a flat and polar structure which is shown in
Fig.~\ref{figBeta}. The potential energy of this structure is 10\,meV/atom higher than the non-polar structure (Fig.~\ref{fig1}). The
vdW, Coulomb, and H-bond energy for the polar monolayer shown in
Fig.~\ref{figBeta} are found to be $E_{coulomb}=$ -0.55\,eV/atom,
$E_{vdW} =$ 2.48\,eV/atom and $ E_{HB} =$ -0.14\,eV/water
respectively. The O-O distance is a=2.88$\pm$0.02\AA~and
the H--O--H angle equals $\theta$=106.6$\pm$+0.1$^o$. We also
performed simulations for the other possible polar structures
proposed by Corsetti \emph{et al.}~\cite{arxiv2015} and found that all of
them have higher energy than the one shown in Fig.~\ref{fig1}.

\begin{figure*}
\begin{center}
\includegraphics[width=0.4\linewidth]{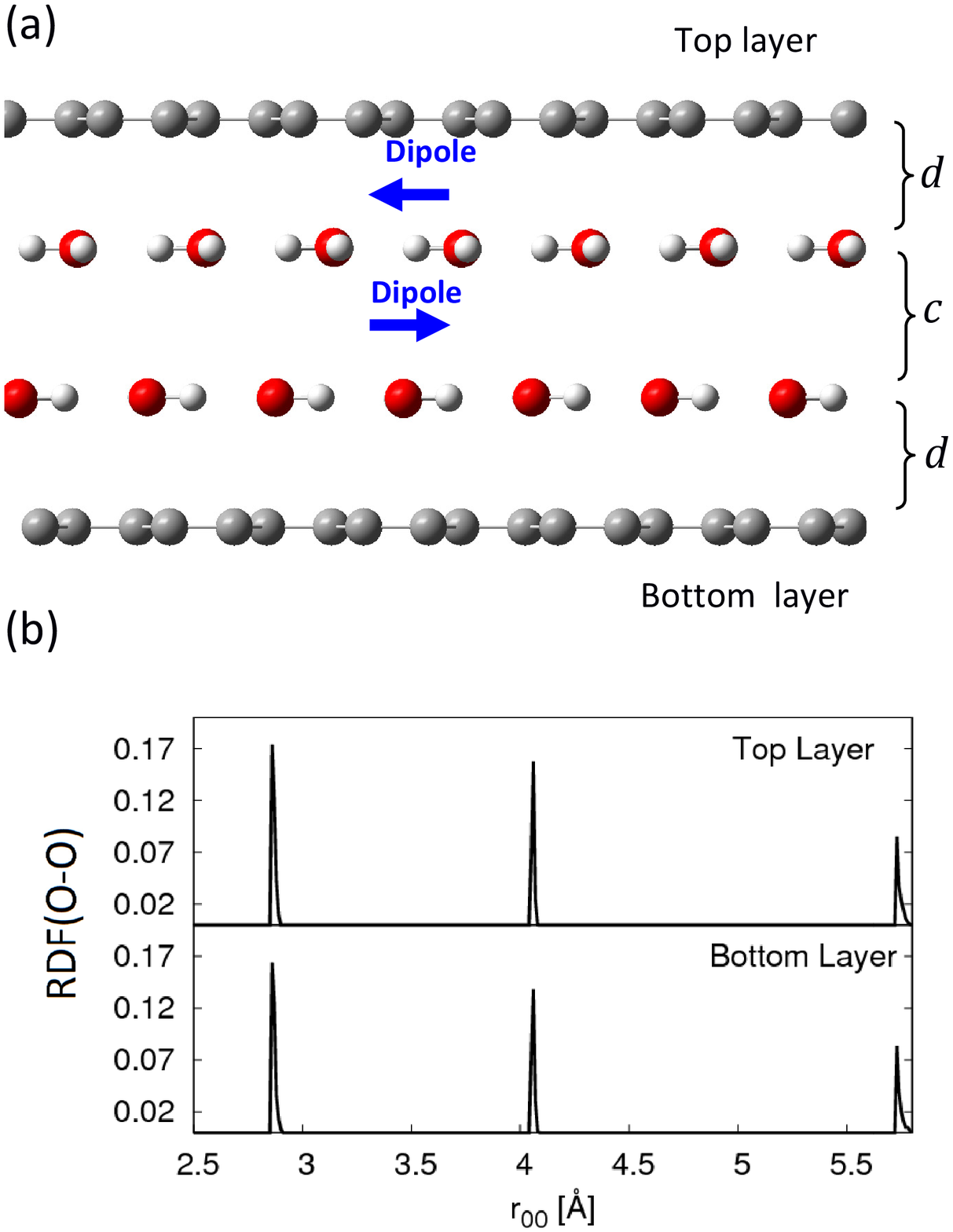}
\includegraphics[width=0.45\linewidth]{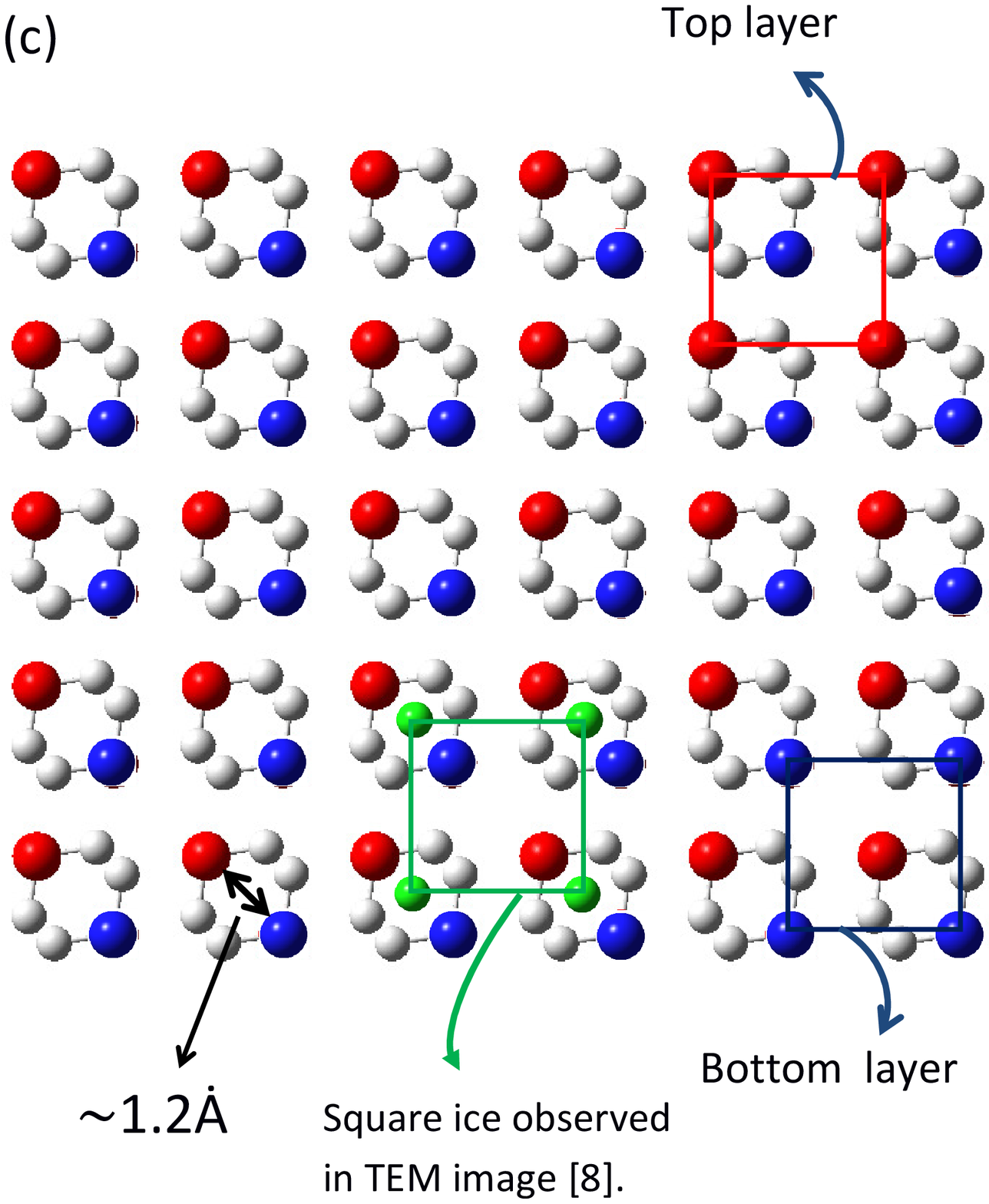}
\caption{\label{fig7} (color online) Relaxed bilayer 
ice between two graphene layers are shown in (a) side and (c) is top view. In (b) the
corresponding radial distribution function for O-O distances is
shown. In the top view (c), the green dots refer to the TEM
experiment~\cite{nat2015}. The in-plane shift between top and bottom
layer is about 1.2\,\AA.}
\end{center}
\end{figure*}

\emph{Bilayer square ice}. Motivated by the experimental work of 
Algara-Siller et al~\cite{nat2015} a bilayer of ice confined between two
rigid graphene layers separated by h=9\AA. The
minimum energy structure is shown in Figs.~\ref{fig7}(a,c).
Surprisingly, also in this case the layers are flat (see
Fig.~\ref{fig7}(a)) and each has a perfect square lattice for the O
atoms (see Fig.~\ref{fig7}(c)). Using the radial distribution
function for each layer we found the lattice constant to be
$a$=2.84$\pm$0.01\AA. The radial distribution function for each ice
layer is presented in Fig.~\ref{fig7}(b) which are identical. The
obtained angle `H--O--H' is found to be $\theta$=106.15$\pm$+0.02$^o$.
Each layer has a net dipole that is in the opposite direction with
respect to the other layer. The latter makes the bilayer of ice
non-polar and non-ferroelectric. The interlayer distance between the
ice layers is found to be c=3.24$\pm$0.01\AA~which is in
disagreement with the SPC/E model which obtained $a\cong c$
~\cite{nat2015}. Our finding for `c' is in the range of vdW adhesion
between two ordinary neutral layers, e.g. the two graphene layers.
The H-bonding, vdW, and the Coulomb energy are -0.13\,eV/water,
2.07\,eV/atom, and -0.86\,eV/atom, respectively. It is interesting
to note that the vdW energy here is lower than that of a single
layer of ice which is due to the extra adhesion between the two ice
layers. Therefore we can estimate the vdW energy stored between the
ice layers as -0.41\,eV/atom. This energetic analysis, gives
important new insights about the physics of confined water between
graphene layers.

The stacking of the two ice layers is not (perfectly) AA stacking,
i.e. the bottom layer has an in-plane shift of about 1.2\AA~(shown
by the arrow in Fig.~\ref{fig7}(c)) with respect to the top layer.
The displacement of the O atoms with respect to each other is due to
the fact that the O-atoms are negatively charged and thus repel each
other. Please note that the SPC/E model~\cite{nat2015} predicts
AB stacking for bilayer ice confined between two graphene layers.
However, the TEM images in the recent experiment~\cite{nat2015},
which are shown by green circles in the bottom right part of
Fig.~\ref{fig7}(c), can be considered as the averaged positions of
oxygen atoms in the top layer (red dots) and bottom layer (blue
dots) of our results. We believe that the blue and red circles in
reality vibrate along the black arrow shown in Fig.~\ref{fig7}(c)
resulting in a time averaged AA-stacking in square ice. In order to investigate the
importance of the interaction between ice and graphene and to
present an independent test, we performed an additional MD
minimization. We minimized the potential energy of monolayer ice
with randomly distributed H-bonds (even out-of-plane) over a single
layer of graphene at an initial distance of 3.0\AA.~The minimum energy
configuration of ice is similar to Fig.~\ref{fig1} with the distance
between graphene and ice of 2.90\AA. This shows that the interaction
between graphene and an ice layer is stronger than the interaction
between two ice layers, which are separated by 3.24\AA~and the
interaction between two graphene layers even if they are at a distance of 9\AA.~

\begin{table*}[htb]
\begin{center}
\begin{tabular}{|c|c|c|c|c|c|c|c|}
\hline
 &   H(\AA)    &    $E_{coulomb}$(eV/atom)  &    $E_{VdW}$(eV/atom)  &   $E_{HB}$(eV/water)    &    $a=r_{oo}=$(\AA)  &   $d$(\AA)            &    $c$(\AA)    \\
\hline
g.ice.g        &   $6.5$ &   $-0.56$         &   $2.43$      &   $-0.16$     &   $2.84\pm0.01$   &   $3.25\pm0.01$   &   $0$     \\
g.ice.ice.g           &   $9$ &   $-0.86$         &   $2.07$      &   $-0.13$     &   $2.84\pm0.01$   &   $2.88\pm0.01$   &   $3.24 \pm0.01$      \\
g.ice.ice.ice.g           &   $11.5$ &   $-0.94$         &   $1.82$      &   $-0.16$     &   $2.89\pm0.02$   &   $2.70\pm0.01$   &   $3.05 \pm0.01$      \\

\hline
\end{tabular}
\caption{Energy contributions for the minimized energy configurations discussed in this chapter. Note that for g.ice.g only the non-polar result is shown, i.e. the one pictured in Fig. \ref{fig1}. } \label{my-label}
\end{center}
\end{table*}

\emph{Thrilayer ice}. Finally, we turn our attention to the
stacking and microscopic structure of confined three layer ice. By
fixing two graphene layers at a distance h=11\AA~ and performing
annealing MD simulations, we found that each layer of the three layer ice
being non-polar with microscopic structure similar to that of
monolayer ice (see Fig.~\ref{fig1}(b)). The ice-graphene distance is
found to be $d=2.70\AA\pm$0.01 and the distance between each of the
ice layers is $c=3.05\AA\pm$0.01. The vdW, Coulomb, and H-bonding
energy are $E_{coulomb}=$ -0.94\,eV/atom, $E_{vdW} =$ 1.82\,eV/atom
and $ E_{HB} =$ -0.16\,eV/water respectively. To have non-polar
multi-layer ice, we conclude that for an odd number of ice layers
each layer has the structure of confined monolayer ice. However a
system with an even number of ice layers confined between graphene
comprises of pairs of layers where each pair has a structure like
bilayer ice (Fig. 5).

\begin{figure}[b]
\includegraphics[width=\linewidth]{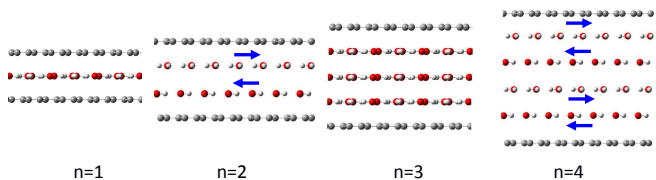}
\caption{\label{figorder} (color online) Multi-layer ice confined
between graphene layers. The arrows indicate the net dipole moment in each layer. }
\end{figure}

\emph{Discussion and conclusions.} In ordinary water the average
distance between oxygen atoms is about 2.82\AA~\cite{chaplin} and
for each water molecule two hydrogen bonds (H-bonds) are randomly
oriented resulting in an irregular network. The H-bonds in liquid
water have a life time of about 1-20 ps. Continuously  bond
formation and bond breaking takes place. The H-bonds are stronger in hexagonal bulk ice
(ordinary ice) though the O-O distance is still 2.82\AA~but each
water molecule takes part in four tetrahedrally-arranged
H-bonds~\cite{chaplin}, i.e. forming a regular network with more
room between the molecules yielding a lower density. Moreover, the
strength and orientation of the H-bonds change when a high pressure
is applied. When water is confined between hydrophobic walls (here
two graphene layers) in the presence of lateral pressure of about 1\,GPa, monolayer ice is formed with the O-O
distance of about 2.83\AA~\cite{nat2015}. Therefore the H-bonds
spacial orientation, strength and their rearrangement is a key
factor which determines the structure of water, its different phases
of ice and confined monolayer, and few layer of bulk     ice.

 In typical ice structures,  e.g. ice $I_h$, the electrostatic attraction between H and O
 atoms dominates the vdW repulsion between the oxygens. However, in a dense monolayer (and few layer) ice,
the electrostatic energy is not large enough to cancel the repulsive
vdW energy, thus, an external pressure is used to keep the system
stable. In fact, the high lateral pressure applied on confined water
at room temperature results in monolayer ice. These conditions
are equivalent to the water phase  close to zero Kelvin (i.e. ice)
at about 1 bar (or zero) which is what we studied. The
multiple configurations of ice at high pressures that meet the rules
of absolute zero amounts to randomness, or in other words, entropy
which is called residence entropy. Therefore the ground state
configuration of water either at zero temperature or high pressures
strongly depends on the experimental procedure and details. This
might be the reason for the observation of several different
structures as reported by different groups for confined ice using
various methods~\cite{nat2015,arxiv2015,ferro}.  The large degree of
freedom for the hydrogen bond strength and its orientation give the
possibility to have several ice structures, either when the system
is subjected to high pressure or is kept close to zero temperatures.
This large degree of freedom is reduced when using the rigid model
in MD simulations which may result in incorrect lattice structure.

 For confined monolayer ice with a flat structure and zero net dipole moment the H-bonds should lie in the same plane. However, flat ice
layers in non-polar bilayer of ice nulceates as a square lattice with an almost AA stacking.
We predict that an odd (even) number of ice layers are stacks of monolayer (bilayer) ice.
We found that the interaction between ice layers is weaker than that between
ice and graphene which results in a shorter distance between ice and
graphene.

{\textit{Acknowledgments}}. \label{agradecimientos} This work was
supported by the ESF-Eurographene project CONGRAN, and the Flemish
Science Foundation (FWO-Vl).

%


\begin{thebibliography}{13}


\bibitem{prl2003} R. Zangi and A. E. Mark, Phys. Rev. Lett.
{\bf91}, 025502 (2003).
\bibitem{jchem2003} R. Zangi and A. E. Mark, J. Chem. Phys {\bf120}, 7123 (2004).


\bibitem{PNAS}J. Bai, C. Austen Angell, and X. Cheng Zeng, PNAS
{\bf107}, 5718 (2010).

\bibitem{ACCOUNT} W-H. Zhao, L. Wang, J. Bai, L-F. Yuan, J. Yang, and X. Cheng
Zeng, Acc. Chem. Res. {\bf47}, 2505 (2014).


\bibitem{arxiv2015}    F. Corsetti, P. Matthews, and E. Artacho, arXiv:1502.03750 (2015).


\bibitem{PCCP}C. Vega, C. C. McBride, E. Sanz E, and JL. Abascal, Phys Chem Chem Phys.  {\bf7}, 1450 (2005).




\bibitem{apl2015} M. Ghosh, L. Pradipkanti, V. Rai, D. K. Satapathy,
P. Vayalamkuzhi, and M. Jaiswa, Applied. Phys. Lett. {\bf106},
241902 (2015)


\bibitem{nat2015}G. Algara-Siller, O. Lehtinen , F. C. Wang , R. R. Nair , U. Kaiser , H. A. Wu
, A. K. Geim, and  I. V. Grigorieva, Nature (London) {\bf519}, 443 (2015).
\bibitem{Novoselov}
K. S. Novoselov, A. K. Geim, S. V. Morozov, D. Jiang, Y. Zhang, S.
V. Dubonos, I. V. Grigorieva, and A. A. Firsov, Science {\bf306},
666 (2004).



\bibitem{prl2012} T. Bjorkman, A. Gulans, A. V. Krasheninnikov, and R. M. Nieminen, Phys. Rev. Lett. {\bf108}, 235502
(2012).

\bibitem{natnano} S. P. Koenig, N. G. Boddeti, M. L. Dunn, and J. S.
Bunch, Nature Nanotechnology {\bf6}, 543 (2011).



%

\bibitem{reax}  A. C. T. van Duin, S. Dasgupta, F. Lorant,  and W. A.
Goddard, J. Phys. Chem. A  {\bf105}, 9396 (2001); A. C. T. van Duin
and J. S. S. Damste, Org. Geochem. {\bf34}, 515 (2003).
\bibitem{lammps} S. Plimpton, J. Comp. Phys. {\bf19}, 117 (1995).
\bibitem{tip2005} J. L. Abascal and C. Vega, J. Chem. Phy. {\bf123},  234505 (2005).



\bibitem{Note1} We emphasize that flat ice with  a
perfect square lattice can not be formed between two graphene layers
because the two electrons on the O atom form two covalent bonds with
the H atoms while the four others make two H-bonds with the neighbor
waters (this is responsible for the asymmetry), i.e. the H-O-H angle
can not be 90$^o$.

\bibitem{private} Private communications with the authors of Ref.
[8].

\bibitem{chaplin}   M. Chaplin,  arXiv:0706.1355 (2007).

\bibitem{prbSTM} M. Neek-Amal, P. Xu , D. Qi , P. M. Thibado , L. O. Nyakiti, V. D. Wheeler
, R. L. Myers-Ward, C. R. Eddy Jr., D.K. Gaskill, and F. M. Peeters,
Phys. Rev. B. {\bf90}, 064101 (2014).


\bibitem{ferro}W.-H. Zhao, J. Bai, L.-F. Yuan, J. Yang, and X. C. Zeng, Chem.
Sci. {\bf5}, 1757 (2014)



\end{thebibliography}
\end{document}